\documentclass[%
 reprint,
 superscriptaddress,
 showpacs,
 amsmath,amssymb,
 aps,
 prd,
 floats,
 nofootinbib,
]{revtex4-1}

\usepackage{dcolumn}
\usepackage{bm}

\usepackage{amsfonts}
\usepackage{latexsym}

\usepackage{float}

\def\bequ{\begin{equation}}
\def\eequ{\end{equation}}
\def\barr{\begin{array}}
\def\earr{\end{array}}

\def\ben{\begin{equation}}
\def\een{\end{equation}}
\def\bena{\begin{eqnarray}}
\def\eena{\end{eqnarray}}

\def\b1{e^0}

\newcommand{\be}{\begin{equation}}
\newcommand{\ee}{\end{equation}}
\def\bea{\begin{eqnarray}}
\def\eea{\end{eqnarray}}

\def\be{\begin{equation}}
\def\ee{\end{equation}}
\def\bea{\begin{eqnarray}}
\def\eea{\end{eqnarray}}

\def\fft#1#2{{\frac{#1}{#2}}}

\def\lesssim{\mathrel{\hbox{\rlap{\hbox{\lower4pt\hbox{$\sim$}}}\hbox{$<$}}}}
\def\gtrsim{\mathrel{\hbox{\rlap{\hbox{\lower4pt\hbox{$\sim$}}}\hbox{$>$}}}}

\begin{document}

\preprint{UPR-1229-T}

\title{Conformal Invariance and  Near-extreme Rotating AdS Black Holes}

\author{Tolga Birkandan}
 \affiliation{Department of Physics and Astronomy, University of Pennsylvania, Philadelphia, PA 19104, USA.}
 \affiliation{Department of Physics, Istanbul Technical University, Istanbul 34469, Turkey.}

\author{Mirjam  Cveti\v c}
 \affiliation{Department of Physics and Astronomy, University of Pennsylvania, Philadelphia, PA 19104, USA.}
 \affiliation{Center for Applied Mathematics and Theoretical Physics, University of Maribor, Maribor, Slovenia}

\begin{abstract}
We obtain  retarded Green's functions  for massless  scalar fields
in the background of near-extreme, near-horizon  rotating charged
black holes of five-dimensional minimal gauged supergravity. The
radial part of the  (separable)  massless Klein-Gordon equation in
such general black hole backgrounds is  Heun's equation,  due to
the  singularity structure associated with the three black hole
horizons. On the other hand, we find the scaling limit for  the
near-extreme, near-horizon  background  where the radial equation
reduces to a hypergeometric equation whose $SL(2,{\bf R})^2$
symmetry  signifies the  underlying two-dimensional conformal
invariance, with  the  two sectors governed by the respective
Frolov-Thorne temperatures.
\end{abstract}


\maketitle

\section{Introduction}

Recently, important new insights into  microscopics   of  extreme  rotating black holes (with zero Hawking temperature) have been obtained \cite{GuicaMU}
 via a  correspondence between these classical gravitational objects
and  the underlying boundary  conformal invariance (Kerr/CFT
correspondence). These new insights are  important for several
reasons; one is that such rotating   black holes  could be
astrophysical: i.e., Kerr black holes  close to extremality could
exist in our universe,  and their microscopics can be addressed
via  Kerr/CFT correspondence \cite{GuicaMU}.

Employing the boundary conformal invariance techniques
\cite{BarnichJY,Barnich}  allows for the microscopic calculation
of extreme rotating black hole entropy which is in agreement with
the Bekenstein-Hawking entropy, as first calculated for extreme
Kerr black holes  in four-dimensions \cite{GuicaMU}. Numerous
follow-up  papers confirmed the agreement for other extreme
rotating black holes. Specifically, this has been verified for  extreme four-dimensional Kerr-Newman (AdS) black holes \cite{Hartman2008}
 and for large classes \cite{ChowDP}   of extreme multicharged rotating
black holes in asymptotically  Minkowski and AdS space-times,  and
in  diverse dimensions.

 Another important  insight into  internal structure of black holes  is obtained by
studying probe particles in these  black hole backgrounds. It was  shown
 in the past that the wave equation  for the massless scalar field  is separable and it has  amazing simplifications
 \cite{CveticUW,CveticXV}, even for general, multicharged rotating black holes  in asymptotic Minkowski space-times \cite{CveticXZ,CveticKV}. In particular,  when certain terms can be neglected the radial part of the wave equations has  an $SL(2,{\bf R})^2$
symmetry, and thus underlying two-dimensional conformal symmetry.
This is the case  for special black hole backgrounds,  such as the
near-extreme limit, with  a microscopic description in terms of
the AdS/CFT correspondence \cite{MaldacenaIX,CveticUW},  and  the
near-extreme rotating limit, with  a recent microscopic calculation
in terms of the Kerr/CFT correspondence \cite{GuicaMU, CveticJN}.
In addition,  this conformal symmetry emerges for  low-energy
scattering   \cite{CveticUW,CastroFD}, as well as in the super-radiant
limit \cite{BredbergPV,CveticJN,Hartman2009}.

On the other hand, for general black hole  backgrounds and for scattering at arbitrary energies the wave equation has no $SL(2,{\bf R})^2$ symmetry,  thus  this is  a stumbling block  for a conformal field theory interpretation in  the general case.
However, a recent proposal  referred to as ``hidden conformal symmetry'' proposes that
the conformal symmetry,  as suggested by  the massless  scalar field wave equation, is still useful
 \cite{CastroFD}, only that it is spontaneously broken.
This approach has been further studied by a number of researchers,
including \cite{Krishnan,Becker2010}. For a review, see
\cite{BredbergHP}.

In a recent development,  in  \cite{CveticLarsen}  an explicit
``subtracted  geometry''  with manifest $SL(2,{\bf R})^2$
conformal invariance was obtained for   general  rotating black
holes in five-dimensional asymptotically flat space-time, by
removing certain ambient flat space-time  terms in an overall warp
factor of the black hole solution. The construction preserves all
near-horizon properties of the black holes, such as the
thermodynamic potentials and the entropy.  The warp factor
subtraction  provides an explicit realization of the conformal
symmetry at the level of the black hole metric.  Furthermore,  a
quantitative relation to  the  standard AdS/CFT correspondence is
obtained by embedding the subtracted black hole geometry  in
auxiliary  six dimensions,  resulting in a long rotating string
with  a fibered $AdS_3\times S^3$
geometry\cite{CveticJA,CveticXH}.

In this paper we advance the study of internal structure and
emergence of conformal invariance for charged rotating black holes
in {\it asymptotically anti-de Sitter space-times}. While the wave
equations for black holes in these backgrounds are again
separable, the radial equation is in general  Heun's equation,
due to more than two horizons of such black holes, and thus the
structure for  such general background, does not exhibit
$SL(2,{\bf R})^2$ conformal invariance. (Note also,  that the study of  microscopics for black holes in  asymptotically anti-de Sitter
space-times remains a complex problem.)  The goal of our analysis
is more modest: We  would like to demonstrate conformal invariance
for the massless scalar field  wave equation in the background of both
extreme {\it and} near-extreme backgrounds of asymptotically AdS black holes.

For concreteness we focus on general charged rotating black  holes
in five-dimensional minimal supergravity \cite{Chong05}. These
black holes are specified by mass, charge, two angular momenta,
and cosmological constant.\footnote{ The uniqueness of the full
solution and the separability of the wave equation for the
massless scalar field  was analyzed by various groups
\cite{Ahmedov:2009ab,Davis:2005ys,Wu2009}. The quasinormal modes
and superradiant instability analysis also exists in the
literature \cite{Aliev:2008yk} for this metric.}

Specifically, we demonstrate that both the extreme and near-extreme limit of these black holes results in
massless scalar wave equations whose solutions  are  hypergeometric functions. (As a prerequisite technical result for  these studies we formulate a precise scaling limit of these  near-extreme, near-horizon  backgrounds.)
We calculate explicitly  the retarded Green's functions both for the extreme and near-extreme  backgrounds, thus generalizing the Green's function results obtained for four-dimensional Kerr backgrounds \cite{Becker2010,BredbergPV}.
The left- and right-moving sectors of boundary conformal theories are governed by a combination of two Frolov-Thorne temperatures \cite{Frolov}, specified along the two combinations of the azimuthal angles.
  These results lend further support for the underlying
conformal field theory description of both extreme and
near-extreme rotating black holes in asymptotically anti-de Sitter
space-times.
 (We expect  that results, analogous to those in this paper,  would be
obtained for other multicharged AdS rotating black holes in four,
five, and possibly other dimensions, e.g.,  black holes of
\cite{Chong06} and references therein.)

The paper is organized  in the following way: In the second section, we
present the radial part of  the massless scalar
field equation  (massless Klein-Gordon equation) with a "natural"
coordinate transformation, and  demonstrate explicitly  that  the pole
structure  is governed by the location of the three horizons and
each residuum  by  the surface gravity at each horizon.  The
resulting equation is Heun's equation.\footnote{The wave
equation thus indicates that all three horizons play a role in the
internal structure of the black hole. A complementary recent
result lends support to this picture, by demonstrating that the
product of areas of all three horizons is quantized  and moduli
independent, i.e. it depends explicitly only on the quantized
charge,  the two  angular momenta and  the  cosmological constant
\cite{Cvetic2010}.}
In the third section, we turn to the extreme black hole background.
 The metric in the extreme case, near-horizon regime  was studied in \cite{ChowDP} and
\cite{Choi:2008he} by taking a specific scaling limit of the
general solution.  Here we analyze the Klein-Gordon equation,
which is a hypergeometric equation with underlying conformal
symmetry, and obtain the explicit retarded  Green's functions
there.  In the fourth section, we address  the near-extreme and
near-horizon limit by  taking a scaling limit where an effective
dimensionless near-extremality parameter is kept explicitly. The
radial equation remains a hypergeometric equation and explicit
retarded Green's functions there.
 In the fifth section, we summarize the results
and comment on future directions. In the Appendix, a further
analysis of the radial equation with a dimensionless coordinate is
presented. We also perform an expansion both for the case of a
small cosmological constant and for a small (non-extremality)
parameter.

\section{The General Metric}
The metric for general black holes in five-dimensional minimal
gauged supergravity was given in Chong et al. \cite{Chong05} and
it can be written in the following form:
\begin{eqnarray}
ds^{2} &=&-\left( dt-\frac{a\sin ^{2}\theta }{\Xi _{a}}\,d\phi
-\frac{b\cos ^{2}\theta }{\Xi _{b}}\,d\psi \right) \notag\\
&&\times\left[ f\left( dt-\frac{a\sin ^{2}\theta }{\Xi
_{a}}\,d\phi \,-\frac{b\cos ^{2}\theta }{\Xi _{b}}\,d\psi
\right) \right.  \notag  \label{gsugrabh} \\
&&\left. +\frac{2Q}{\Sigma }\left( \frac{b\sin ^{2}\theta }{\Xi _{a}}\,d\phi
+\frac{a\cos ^{2}\theta }{\Xi _{b}}\,d\psi \right) \right] \notag\\&&+\,\Sigma \left(
\frac{r^{2}dr^{2}}{\Delta _{r}}+\frac{d\theta ^{\,2}}{~\Delta _{\theta }}%
\right)
+\,\frac{\Delta _{\theta }\sin ^{2}\theta }{\Sigma }\left( a\,dt-\frac{%
r^{2}+a^{2}}{\Xi _{a}}\,d\phi \right) ^{2} \notag \\&&+\,\frac{\Delta _{\theta }\cos
^{2}\theta }{\Sigma }\left( b\,dt-\frac{r^{2}+b^{2}}{\Xi _{b}}\,d\psi
\right) ^{2}+\,\frac{1+r^{2}\,G}{r^{2}\Sigma } \notag \\&& \times\left(ab\,dt-\frac{b(r^{2}+a^{2})\sin
^{2}\theta }{\Xi _{a}}\,d\phi-\,\frac{a(r^{2}+b^{2})\cos ^{2}\theta }{\Xi_{b}}\,d\psi \right)^{2},
\end{eqnarray}
where,
\begin{eqnarray}
f &=&\frac{\Delta _{r}-2ab\,q-q^{2}}{r^{2}\Sigma }+\frac{q^{2}}{\Sigma ^{2}}%
,~~\notag\\\Sigma &=&r^{2}+a^{2}\cos ^{2}\theta +b^{2}\sin ^{2}\theta ,  \notag \\
~\Xi _{a} &=&1-a^{2}G\,\,,~~~~\Xi _{b}=1-b^{2}G,  \notag \\
\Delta _{r} &=&\left( r^{2}+a^{2}\right) \left( r^{2}+b^{2}\right) \left(
1+r^{2}G\right) \notag\\&&+2ab\,q+q^{2}-2mr^{2},  \notag \\
\Delta _{\theta } &=&1-G(a^{2}\cos ^{2}\theta -b^{2}\,\sin
^{2}\theta ).~~~~~
\end{eqnarray}
Here, we should note that we define the parameter $G\equiv g^{2}$,
related to the cosmological constant $\Lambda=-6G$. The ``bare''
black hole parameters $m,q,a,b$, and $G$ specify the physical
parameters mass $M$, charge $Q$ and two angular momenta in the
following way:
\bea
M &=& \fft{m\pi(2\Xi_a+2\Xi_b - \Xi_a\, \Xi_b) + 2\pi q a b
G(\Xi_a+\Xi_b)}{4 \Xi_a^2\, \Xi_b^2}\,,\\
 Q &=& \fft{\sqrt3\, \pi q}{4 \Xi_a\, \Xi_b}\,,\\
J_a &=& \fft{\pi[2am + qb(1+a^2 G) ]}{4 \Xi_a^2\, \Xi_b}\,,\\
J_b &=& \fft{\pi[2bm + qa(1+b^2 G) ]}{4 \Xi_b^2\, \Xi_a}\,.
\eea
The Klein-Gordon equation for a massless scalar field is given by,
\begin{equation}
\frac{1}{\sqrt{-g}}\partial _{\mu }\left( \sqrt{-g}g^{\mu \nu }\partial
_{\nu }\Phi \right) =0,
\end{equation}
where $g$ is the determinant of the metric. The scalar field Ansatz can be
written as,
\begin{equation}
\Phi =e^{-i\omega t+im_{1}\phi +im_{2}\psi }\tilde{R}(r)F(\theta ).
\end{equation}
For the sake of simplicity,  the angular eigenvalues will be set to zero ($%
m_{1}=m_{2}=0$) in this section. (For non-zero $m_{1,2}$ parameters
the wave equation exhibits analogous pole structure, but with the residua are corrected accordingly. For general asymptotically flat black holes this has been done explicitly in \cite{CveticUW}.)

The radial equation then reads,
\begin{equation}
\begin{split}
\frac{\Delta _{r}}{r}\frac{d}{dr} \left( \frac{\Delta
_{r}}{r}\frac{d\tilde{R} }{dr}\right) + \bigg\{
\frac{[(r^{2}+a^{2})(r^{2}+b^{2})+abq]^{2} \omega ^{2}}{r^{2}}
\\ -\Delta_{r}\left( c_{0}+\frac{a^{2}b^{2}}{r^{2}}\omega^{2}\right) \bigg\} \tilde{R}=0,
\end{split}
\end{equation}
where $c_{0}$ is the separation constant, which is the eigenvalue
of the angular equation, displayed in  the next subsection.

As the horizon equation $\Delta _{r}$ is quadratic in $r$, a coordinate
transformation $u=r^{2}$  renders the equation in  the following form:
\begin{equation}
\begin{split}
\frac{d}{du}\left( \Delta _{u}\frac{dR}{du}\right)
+\frac{1}{4}\bigg\{ \bigg[
\frac{[(u+a^{2})(u+b^{2})+abq]^{2}}{u\Delta _{u}}
\\-\frac{a^{2}b^{2}}{u} \bigg]
\omega ^{2}-c_{0} \bigg\} R=0,
\end{split}
\end{equation}
where $R\equiv R(u)$ and $\Delta_u\equiv  \Delta_u(u)$. From now on, the dependence of the radial function
will not be updated and the radial function will be denoted as $R$ for each
case.

We cast the equation in a  form,  which exhibits the singularity structure
 more transparently,
\begin{equation}
\begin{split}
\frac{d}{du}\left( \Delta _{u}\frac{dR}{du}\right)
+\frac{1}{4}\bigg\{ \bigg[
\frac{n_{1}}{\kappa _{1}^{2}(u-u_{1})}+\frac{n_{2}}{\kappa _{2}^{2}(u-u_{2})}%
\\+\frac{n_{3}}{\kappa _{3}^{2}(u-u_{3})}+Gn_{4}\bigg] \omega
^{2}-c_{0}\bigg\} R=0,\label{radial}\,
\end{split}
\end{equation}
where the $\kappa _{i}$'s are the surface gravities associated with
 each horizon located at  $u_{i}$.  The  horizon  coordinates $u_i$ can be determined in terms of the bare black hole parameters via the horizon equation:
\begin{equation}
\Delta _{u}\equiv G(u-u_{1})(u-u_{2})(u-u_{3}),
\end{equation}
which provides the following relations:
\begin{eqnarray}
\overset{3}{\underset{i=1}{\sum }}u_{i} &=&-\left( \frac{1}{G}%
+a^{2}+b^{2}\right) , \\
\underset{j<i=1}{\overset{3}{\sum }}u_{i}u_{j} &=&\frac{1}{G}\left(
a^{2}+b^{2}+a^{2}b^{2}G-2m\right) , \\
\overset{3}{\underset{i=1}{\prod }}u_{i} &=&-\frac{(q+ab)^{2}}{G}.
\end{eqnarray}
The surface gravities are given by \cite{Chong05},
\bea
\kappa _{i}=\frac{u_{i}^{2}\left[ 1+G\left( 2u_{i}+a^{2}+b^{2}\right) \right]
-(ab+q)^{2}}{\sqrt{u_{i}}\left[ (u_{i}+a^{2})(u_{i}+b^{2})+abq\right] }\text{
\ }\,
\eea
where $i=1, 2, 3$ and they can be rewritten as,
\begin{eqnarray}
\kappa _{1} &=&\frac{G(u_{1}-u_{3})(u_{1}-u_{2})\sqrt{u_{1}}}{%
(u_{1}+a^{2})(u_{1}+b^{2})+abq}, \\
\kappa _{2} &=&\frac{G(u_{2}-u_{3})(u_{1}-u_{2})\sqrt{u_{2}}}{%
(u_{2}+a^{2})(u_{2}+b^{2})+abq}, \\
\kappa _{3} &=&\frac{G(u_{2}-u_{3})(u_{1}-u_{3})\sqrt{u_{3}}}{%
(u_{3}+a^{2})(u_{3}+b^{2})+abq}.
\end{eqnarray}
The $n_{i}$ coefficients can be found after some algebra as,
\begin{eqnarray}
n_{1} &=&G(u_{1}-u_{2})(u_{1}-u_{3}), \\
n_{2} &=&-G(u_{1}-u_{2})(u_{2}-u_{3}), \\
n_{3} &=&G(u_{1}-u_{3})(u_{2}-u_{3}), \\
n_{4} &=&\frac{1}{G^{2}}.
\end{eqnarray}
It is important to note that  the radial equation (\ref{radial})
has a pole structure, which is  determined by the  location of the
three horizons and the residuum of the pole at  each horizon is
inversely  proportional for the square of the surface gravity
there. This result signifies the role that all three horizons have
in determining the internal structure of such black holes. On the
other hand, the resulting  equation is Heun's equation and it does
not have a compact form of a solution, not even in the case of the
low-energy $\omega$ of the  scalar field. In the Appendix we
further analyze the radial wave equation in terms of a
dimensionless coordinate; we also perform a small $G$ expansion,
and an expansion in the near-extreme limit, suitable for the
analysis in the subsequent sections.

\subsection{The angular equation}
Here we comment on the angular equation, which is for the general metric and taking $m_{1}=m_{2}=0$ it is of the form,
\begin{equation}
\begin{split}
&y^{4}Y^{2}\frac{d^{2}F}{dy^{2}}+y^{3}Y\left( Y+y\frac{dY}{dy}\right) \frac{dF%
}{dy}
\\&-\big[ (a^{2}b^{2}\omega
^{2}+c_{0}y^{2})y^{2}Y
\\&+(a-y)^{2}(a+y)^{2}(b-y)^{2}(b+y)^{2}\omega ^{2}\big]
F=0,\label{ang0}
\end{split}
\end{equation}
where,
\begin{eqnarray}
y &=&\sqrt{a^{2}\cos ^{2}\theta +b^{2}\sin ^{2}\theta }, \\
Y &=&-\frac{(1-Gy^{2})(a^{2}-y^{2})(b^{2}-y^{2})}{y^{2}}.
\end{eqnarray}
Davis et al. have studied the angular equation without taking $%
m_{1}=m_{2}=0$ and found that the solution can be given by
general Heun's functions \cite{Davis:2005ys}. The structure of the
equation does not change for $m_{1}=m_{2}=0$ and the solutions can
still be given in terms of general Heun's functions
\cite{ronvbk,slavbk,Birkandan:2006ac,hortacsuheun}. The
angular equation in another parametrization was studied in \cite%
{Aliev:2008yk} as a Sturm-Liouville problem and the eigenfunctions  are
the AdS modified spheroidal functions in five dimensions.

The angular part of the wave equation  is independent of the non-extremality parameter, and thus the extreme and near-extreme cases, studied in the subsequent sections with general $m_{1,2}$,  result in the same form of the angular equation as the general background.
Therefore,  for the sake of completeness we write below the angular equation  with non-zero $m_{1,2}$ in the notation suitable for our subsequent analysis (in terms of $u_1, G, a, b, q$), as,
\begin{equation}
\begin{split}
&{Y}\frac{d^{2}F}{d{y}^{2}}+{\frac{1}{y}\left( y{\frac{dY}{dy}}
+Y\right) }\frac{dF}{dy} +\\& \bigg( \frac{X}{\left[ \left( {b}
^{2}+u_{1}\right) ({a}^{2}+u_{1})+abq\right] ^{2}{y}^{4}Y}
\\&-2G(u_{1}-u_{3})c_{1,2}\bigg) F=0,\label{ang}
\end{split}
\end{equation}
where,
\begin{widetext}
\begin{eqnarray}
X &=&-\left\{ m_{2}\,b\left( u_{1}+{y}^{2}\right) \left( -1+{b}^{2}G\right) {%
a}^{4}\right.   \notag \\
&&+\left[ Gm_{1}\,\left( u_{1}+{y}^{2}\right) {b}^{4}+G\left( {y}%
^{2}m_{2}\,q+m_{1}u_{1}^{2}-{y}^{4}m_{1}\right) {b}^{2}-{y}^{2}\left(
u_{1}\,Gm_{1}\,{y}^{2}+Gm_{1}u_{1}^{2}+m_{2}\,q\right) \right] {a}^{3}
\notag \\
&&+\left[ G\left( u_{1}^{2}m_{2}-{y}^{4}m_{2}+{y}^{2}m_{1}\,q\right) {b}%
^{2}+\left( -Gm_{1}\,q+m_{2}\right) {y}^{4}-u_{1}^{2}m_{2}\right] b{a}^{2}
\notag \\
&&+\left[ -m_{1}\,\left( u_{1}+{y}^{2}\right) {b}^{4}+\left[ \left(
-Gm_{2}\,q+m_{1}\right) {y}^{4}-m_{1}\,u_{1}^{2}\right] {b}^{2}+{y}^{2}\left[
\left( m_{1}\,u_{1}+m_{2}\,q\right) {y}^{2}+m_{1}\,u_{1}^{2}\right] \right] a
\notag \\
&&\left. -b{y}^{2}\left[ \left( qm_{1}+u_{1}\,m_{2}\,G{y}^{2}+Gm_{2}%
\,u_{1}^{2}\right) {b}^{2}+\left( -u_{1}\,m_{2}-qm_{1}\right) {y}%
^{2}-u_{1}^{2}m_{2}\right] \right\} ^{2} \\
&&-\left( u_{1}+{y}^{2}\right) u_{1}\,{y}^{2}\left\{ \left[ Gm_{1}\,b{a}%
^{2}+m_{2}\,\left( -1+{b}^{2}G\right) a-m_{1}\,b\right] u_{1}+m_{2}\left( -1+%
{b}^{2}G\right) {a}^{3}+Gm_{1}\,{b}^{3}{a}^{2}-m_{1}\,{b}^{3}\right\} ^{2}Y,
\notag
\end{eqnarray}
\end{widetext}
and $c_{1,2}$  are separation constants, suitable for the
parametrization in  respective extreme and near-extreme cases.
Note that $c_0$ separation constant in Eq.(\ref{ang0}) is related
to $c_{1,2}$ via a $2G(u_1-u_3)$ factor.

The further analysis of the angular equation is beyond the scope of our
study.

\section{Radial equation and Green's function for the extreme near-horizon
limit}

The extreme near-horizon
scaling limit of the original
metric was derived in Chow et al. \cite{ChowDP}. The extreme limit implies,
\begin{equation}
u_{2}=u_{1},
\end{equation}
and the  scaling transformation  of Chow et al. \cite{ChowDP}  on coordinates takes  the form,
\begin{eqnarray}
u &=&u_{1}(1+\eta \rho)^{2}, \notag \\
\hat{t} &=&\beta t, \notag\\
\hat{\phi}_{1} &=&\phi _{1}+\Omega _{1}\hat{t}, \notag\\
\hat{\phi}_{2} &=&\phi _{2}+\Omega _{2}\hat{t},
\label{coord}
\end{eqnarray}
where,
\begin{eqnarray}
\beta &=&\frac{1}{4\pi u_{1}\left( \frac{\partial T_{H}}{\partial u_{1}}%
\right) \eta }, \\
\Omega _{1} &=&\frac{\Xi _{a}(au_{1}+ab^{2}+qb)}{%
(u_{1}+a^{2})(u_{1}+b^{2})+qab}, \\
\Omega _{2} &=&\frac{\Xi _{b}(bu_{1}+a^{2}b+qa)}{%
(u_{1}+a^{2})(u_{1}+b^{2})+qab}.
\end{eqnarray}
Here, $\Omega _{1}$ and $\Omega _{2}$ are the angular velocities
and $T_{H}$ is the Hawking temperature at the horizon $u_{1}$
given by Chow et al. \cite{ChowDP}.

The Vielbeine are obtained by taking  the scaling limit $\eta \rightarrow 0$ and are given explicitly on \cite{ChowDP}. Note
that this specific scaling limit  of the extreme solution
``magnifies'' the near-horizon region.

The
solution  Ansatz of the massless Klein-Gordon equation is of  the
following form,
\begin{equation}
\Phi =e^{-i\omega t+im_{1}\phi _{1}+im_{2}\phi _{2}}R(\rho )F(y).
\end{equation}
The angular part $F(y)$ satisfies Heun's equation (\ref{ang}). On
the other hand, the radial equation for the extreme near-horizon
scaling limit  is given by,
\begin{widetext}
\begin{equation}
\frac{d}{d\rho }\left( {{\rho }^{2}{\frac{dR}{d{\rho }}}}\right) {+}%
\left\{ {\frac{[(m_{1}T_{2}+m_{2}T_{1})^{2}-2\,c_{1}\,{\pi }%
^{2}T_{1}^{2}T_{2}^{2}]{\rho }^{2}+2\,\pi \,T_{1}\,T_{2}\,\left(
m_{1}T_{2}+m_{2}T_{1}\right) \omega \rho +{\pi }^{2}T_{1}^{2}T_{2}^{2}{%
\omega }^{2}}{4{\rho }^{2}{\pi }^{2}T_{1}^{2}T_{2}^{2}}}\right\} {R}{%
=0,}
\end{equation}
\end{widetext}
where $c_{1}$ is the separation constant. $T_{1}$ and $T_{2}$ are
the Frolov-Thorne temperatures given by Chow et al. \cite{ChowDP},
\begin{eqnarray}
T_{1} &=&\frac{G\sqrt{u_{1}}(u_{1}-u_{3})\left[ \left(
a^{2}+u_{1}\right) \left( b^{2}+u_{1}\right) +abq\right] }{\pi \Xi
_{a}\left[ a\left(
b^{2}+u_{1}\right) ^{2}+bq\left( b^{2}+2u_{1}\right) \right] },  \notag \\
T_{2} &=&\frac{G\sqrt{u_{1}}(u_{1}-u_{3})\left[ \left(
a^{2}+u_{1}\right) \left( b^{2}+u_{1}\right) +abq\right] }{\pi \Xi
_{b}\left[ b\left( a^{2}+u_{1}\right) ^{2}+aq\left(
a^{2}+2u_{1}\right) \right] },
\end{eqnarray}
where we used a suitable notation for our purposes.

In the extreme limit, as $T_{H}=0$ and $G$ can be related to $u_1, a,b,q$ as
\begin{equation}
G=\frac{(ab+q)^{2}-u_{1}^{2}}{u_{1}^{2}(a^{2}+b^{2}+2u_{1})}.
\end{equation}
$G$ enters the equations via $\Xi _{a}=1-a^{2}G$ and $\Xi _{b}=1-b^{2}G$, and also in value of $u_3$ as,
\begin{equation}
u_3=-\frac{(q+ab)^2}{G u_1^2 }.
\end{equation}
Note that the radial equation is  written explicitly for non-zero
values of $m_{1,2}$. Its solutions are special hypergeometric
functions, i.e.,  Whittaker  functions.  Thus, the equation
possesses the conformal $SL(2,{\bf R})^2$ symmetry. The explicit
form of the radial solution can be written as a linear combination
of two Whittaker functions $M$ and $W$,
\bea
R &=& \mathit{C}_{\mathit{1}}^{\prime }\,M{\bigg( {\frac{-\,iH_{1} }{2\,H_{2}}}
,\sqrt{\frac{1}{2}\,\left( c_{1}+\frac{1}{2}\right) -\frac{H_{1}^{2}}{4H_{2}^{2}}},\,{\frac{i\omega }{\rho }}\bigg) }
\notag \\&+&\mathit{C}_{\mathit{2}}^{\prime }\,W{\bigg( {\frac{-\,iH_{1} }{2\,H_{2}}}
,\sqrt{\frac{1}{2}\,\left( c_{1}+\frac{1}{2}\right) -\frac{H_{1}^{2}}{4H_{2}^{2}}},\,{\frac{i\omega }{\rho }}\bigg) ,}
\eea
where $\mathit{C}_{\mathit{1}}^{\prime }$ and $\mathit{C}_{\mathit{2}%
}^{\prime }$ are constants and $c_{{1}}$ is the separation constant. Here we
defined,
\begin{eqnarray}
H_{1} &=&m_{1}T_{2}+m_{2}T_{1}, \\
H_{2} &=&\pi \,T_{1}T_{2}.
\end{eqnarray}
The retarded Green's function can be obtained, analogous to  four-dimensional Kerr backgrounds \cite{Becker2010},
as,
\begin{equation}
G_{R}\sim \frac{{\mathcal{B}}(m_1,m_2,\omega )}{{\mathcal{A}}(m_1,m_2,\omega )},
\end{equation}
where the coefficients ${\cal {A}}$ and ${\cal B}$ are determined by the   asymptotic expansion of the radial solution,  given in the form,
\begin{equation}
R _{m_1,m_2,\omega }^{in}\sim N\left( \mathcal{A}\rho^{-\frac{1}{2}+\beta }+%
\mathcal{B}\rho^{-\frac{1}{2}-\beta }+\text{higher order terms}\right) .
\end{equation}
where,
\begin{equation}
\beta =\sqrt{\frac{1}{2}\left( c_{1}+\frac{1}{2}\right) -\frac{H_{1}^{2}}{%
4H_{2}^{2}}}-\frac{iH_{1}}{2H_{2}}.
\end{equation}%
The Green's function can be calculated accordingly (by expanding  Whittaker functions),
\begin{widetext}
\begin{equation}
G_{R}\sim \frac{(i\omega )^{\sqrt{2c_{1}-\frac{H_{1}^{2}}{H_{2}^{2}}+1}%
}\Gamma \left[ -\sqrt{-\frac{H_{1}^{2}}{H_{2}^{2}}+2c_{1}+1}\right] \Gamma %
\left[ \frac{1}{2}\left( \frac{iH_{1}}{H_{2}}+\sqrt{-\frac{H_{1}^{2}}{%
H_{2}^{2}}+2c_{1}+1}+1\right) \right] }{\Gamma \left[ \sqrt{-\frac{H_{1}^{2}%
}{H_{2}^{2}}+2c_{1}+1}\right] \Gamma \left[ \frac{1}{2}\left( \frac{iH_{1}}{%
H_{2}}-\sqrt{-\frac{H_{1}^{2}}{H_{2}^{2}}+2c_{1}+1}+1\right) \right] }.
\end{equation}
\end{widetext}
It is interesting to observe that  both Frolov-Thorne temperatures
enter this Green's function, as they govern the two sectors of
underlying  the two-dimensional conformal field theory (CFT).
These Green's functions can be further employed in the AdS/CFT
context for the calculation of Lorentzian signature correlation
functions for  boundary field theory operators \cite{Son}. The
retarded Green's functions for four-dimensional  Kerr backgrounds
were employed by Becker et al. \cite{Becker2010} for such
two-point and three-point correlation functions. The explicit form
of  retarded Green's function obtained in this section provides a
starting point for quantitative studies of the boundary field
theory correlators  for the extreme charged AdS backgrounds.

\section{Radial equation and Green's function for the near-extreme
near-horizon limit}

In this section we derive the radial part of the wave equation in the near-extreme, near-horizon limit.  For that purpose, we proceed with
the expansion around the extreme limit $u_1=u_2$, i.e. when the two horizons coincide.
Thus, a deviation from extremality can be defined as,
\begin{equation}
u_{2}=u_{1}(1+p\eta ),
\end{equation}
where the small parameter $\eta\ll 1$. As we shall  shortly see,
$p$ can be regarded as a (dimensionless)  measure  of  the
``effective near-extremality'' and it does not have to be  small.
Furthermore, we introduce the scaling transformation of the metric
coordinates  (\ref{coord}). Now, the scaling limit $\eta \to 0$
signifies   a ``magnification'' of near-extreme, near-horizon
geometry.

In this limit the Vielbeine of the original general metric,  which were given in \cite{ChowDP}, take the form,
\bea
&&e^{0}=\sqrt{\frac{u_{1}+y^{2}}{2G(u_{1}-u_{3})}}\sqrt{\rho(2\rho-p)}dt, \\
&&e^{1}=\sqrt{\frac{u_{1}+y^{2}}{2G(u_{1}-u_{3})}}\frac{d\rho}{\sqrt{\rho(2\rho-p)}}, \\
&&e^{2}=\sqrt{\frac{Y}{u_{1}+y^{2}}}\left(\frac{a\tilde{e}_{1}\left(
a^{2}+u_{1}\right) }{\left( a^{2}-b^{2}\right) ~\Xi _{a}}+\frac{b\tilde{e}
_{2}\left( b^{2}+u_{1}\right) }{\left( b^{2}-a^{2}\right) \Xi _{b}}\right) ,\\
&&e^{3}=\sqrt{\frac{u_{1}+y^{2}}{Y}}dy, \\
&&e^{4}=\frac{ab}{\sqrt{u_{1}}y}\bigg( \frac{\tilde{e}_{1}\left(
a^{2}-y^{2}\right) \left( b\left( a^{2}+u_{1}\right) \left(
u_{1}+y^{2}\right) +aqy^{2}\right) }{ab\left( a^{2}-b^{2}\right) \left(
u_{1}+y^{2}\right) ~\Xi _{a}} \notag
\\&&+\frac{\tilde{e}_{2}\left( b^{2}-y^{2}\right)
\left( a\left( b^{2}+u_{1}\right) \left( u_{1}+y^{2}\right) +bqy^{2}\right)
}{ab\left( b^{2}-a^{2}\right) \left( u_{1}+y^{2}\right) \Xi _{b}}\bigg),
\eea
where,
\begin{eqnarray}
Y &=&-\frac{(1-Gy^{2})(a^{2}-y^{2})(b^{2}-y^{2})}{y^{2}},  \notag \\
\tilde{e}_{1} &=&-\left( d\phi _{1}+\frac{\rho}{\pi
T_{1}}dt\right) ,\\
\tilde{e}_{2}&=&-\left( d\phi _{2}+\frac{\rho}{\pi T_{2}}%
dt\right) ,  \notag
\end{eqnarray}
and, $T_{i}$'s are the Frolov-Thorne temperatures. This  new
explicit metric,  obtained above by taking  the  near-extreme,
near-horizon scaling limit, is now  parameterized by the effective
near-extremality  parameter $p$. Of course, as $p\rightarrow  0$,
one obtains the extreme near-horizon  metric  \cite{ChowDP}, which
was  employed in the previous section.

The Ansatz for the scalar function is chosen as,
\begin{equation}
\Phi =e^{-i\omega t+im_{1}\phi _{1}+im_{2}\phi
_{2}}R(\rho)F(y).
\end{equation}
The radial equation is given as,
\begin{widetext}
\begin{equation}
\frac{d}{d\rho}\left[ {{\rho}\left( p-2\,\rho\right) {\frac{dR}{d{\rho}}}}\right] +%
\frac{\left[ c_{2}\pi ^{2}T_{1}^{2}T_{2}^{2}(p-2\rho)\rho+\left(
m_{1}T_{2}+m_{2}T_{1}\right) ^{2}\rho^{2}+2\pi
T_{1}T_{2}(m_{1}T_{2}+m_{2}T_{1})\omega \rho+{\pi }^{2}T_{1}^{2}T_{2}^{2}{%
\omega }^{2}\right] }{{\pi }^{2}T_{1}^{2}T_{2}^{2}\left( p-2\,\rho\right) \rho}%
R=0,
\end{equation}
and as $p\rightarrow 0$ this equation agrees exactly with the
extreme case of the previous section. This equation is also a
hypergeometric equation, and thus possesses the $SL(2,{\bf R})^2$
invariance associated with the conformal symmetry.  The solution
can be written in terms of  hypergeometric functions as,
\begin{eqnarray}
R &=&\left( -p+2\,\rho\right) ^{{\frac{\,i\left[ H_{1} p+2\,\omega H_{2}\right] }{2H_{2}\,p}}}  \notag \\
&&\times \left\{ {\rho}^{{\frac{-i\omega
}{p}}}\mathit{C}_{\mathit{1}}^{\prime \prime }\,{\
_{2}F_{1}\,}\left[ {{\frac{H_{2}-\sqrt{\left(
1+2c_{2}\right) H_{2}^{2}-H_{1}^{2}}%
+iH_{1}}{2H_{2}}}}\right. \right.
\left. {,{\frac{H_{2}+\sqrt{\left( 1+2c_{2}\right) H_{2}^{2}-H_{1}^{2}}+iH_{1}%
}{2H_{2}}};\,{\frac{p-2\,i\omega
}{p}};\,{\frac{2\rho}{p}}}\right]
\notag \\
&&+{\rho}^{{\frac{i\omega }{p}}}\mathit{C}_{\mathit{2}}^{\prime
\prime }\,{\ _{2}F_{1}\,}\left[ {{\frac{4\,i\omega H_{2}+\left( \sqrt{\left(
1+2c_{2}\right) H_{2}^{2}-H_{1}^{2}}%
\right) p+\left[ iH_{1}+H_{2}\right] p}{2H_{2}\,p}}}\right. \\
&& \hspace{1in} \left. \left. {,\,{\frac{4\,i\omega H_{2}-\left( \sqrt{\left(
1+2c_{2}\right) H_{2}^{2}-H_{1}^{2}}%
\right) p+\left[ iH_{1}+H_{2}\right]
p}{2H_{2}\,p}};\,{\frac{p+2\,i\omega
}{p}};\,{\frac{2\rho}{p}}}\right] \right\} .  \notag
\end{eqnarray}
The retarded  Green's function  can furthermore be obtained by determining the ratio of the coefficients in the asymptotic expansion of the above solution as,
\begin{eqnarray}
G_{R} &\sim &2^{-\sqrt{2c_{2}-\frac{H_{1}^{2}}{H_{2}^{2}}+1}}\left( -\frac{1%
}{p}\right) ^{-\sqrt{2c_{2}-\frac{H_{1}^{2}}{H_{2}^{2}}+1}}   \\
&&\times \frac{\Gamma \left[ -\sqrt{-\frac{H_{1}^{2}}{H_{2}^{2}}+2c_{2}+1}%
\right] \Gamma \left[ \frac{1}{2}\left( \frac{iH_{1}}{H_{2}}+\sqrt{-\frac{%
H_{1}^{2}}{H_{2}^{2}}+2c_{2}+1}+1\right) \right] \Gamma \left[ \frac{1}{2}%
\left( -\frac{4i\omega }{p}+\sqrt{-\frac{H_{1}^{2}}{H_{2}^{2}}+2c_{2}+1}-%
\frac{iH_{1}}{H_{2}}+1\right) \right] }{\Gamma \left[ \sqrt{-\frac{H_{1}^{2}%
}{H_{2}^{2}}+2c_{2}+1}\right] \Gamma \left[ \frac{1}{2}\left( \frac{iH_{1}}{%
H_{2}}-\sqrt{-\frac{H_{1}^{2}}{H_{2}^{2}}+2c_{2}+1}+1\right) \right] \Gamma %
\left[ \frac{1}{2}\left( -\frac{4i\omega }{p}-\sqrt{-\frac{H_{1}^{2}}{%
H_{2}^{2}}+2c_{2}+1}-\frac{iH_{1}}{H_{2}}+1\right) \right] }. \notag
\end{eqnarray}
\end{widetext}
Note again that the retarded Green's function, while a bit more complex, carries an explicit dependence on both Frolov-Thorne temperatures as well as the effective non-extremality parameter $p$. Again this result is a starting point in the calculation of the Lorentzian signature correlation functions of the boundary conformal field theory operators \cite{Son} , analogous to calculations for the four-dimensional Kerr backgrounds in \cite{Becker2010}.
As $p\rightarrow 0$, the $\frac{4i\omega }{p}$ term dominates both in the
numerator and in the denominator. Thus, these gamma functions cancel each
other as the additional terms would become insignificant. Then, up to the
factors in front of the gamma functions, we have the result we obtained in
the extreme case.

\section{Conclusions}

The aim of this paper was to  gain insight into the internal
structure of rotating charged black holes in the asymptotically
anti-de Sitter space-times. For this purpose  we presented  a
comprehensive study of the massless scalar wave equation in the
black hole background of the five dimensional minimal gauged
supergravity. We focused on the radial part of such a separable
wave equation, as the angular part was addressed in the past in
\cite{Davis:2005ys,Aliev:2008yk} in terms of general Heun's
functions \cite{ronvbk,slavbk,Birkandan:2006ac,hortacsuheun}.

For the general black hole background the radial equation is
Heun's equation, which can be cast, after a suitable coordinate
transformation in a form  whose structure  is governed by the
three poles associated with the locations of the three horizons
and each residuum is inversely proportional to the square of the
surface gravity there. The result signifies an importance of {\it
all three horizons} when probing the  internal structure of such
black holes. It  also complements a recent result of
\cite{Cvetic2010}, where it was shown that  the area product of
the  three horizons is quantized,  an indication that all three
horizons may play a role in microscopics  of such general black
holes.

We further focused on the study of the extreme and  near-extreme
limits of such black holes and the explicit solutions of the
massless scalar wave equation in the near-horizon regime. For the
extreme case such a scaling limit was studied in \cite{ChowDP}. We
find that the radial equation in this scaling limit is an equation
for special hypergeometric (Whittaker) functions, and thus
signifies the $SL(2, {\rm R})^2$ conformal invariance.  The
explicit retarded Green's function has a dependence on the two
Frolov-Thorne temperatures, associated with the two sectors of the
two-dimensional conformal theory.

In order to address the near-extreme near-horizon regime, we
introduce a scaling limit, which corresponds to the same scaling
transformations of the coordinates as in the extreme case;
however, it  also introduces an additional ``effective
near-extremality' parameter'' which remains finite and
dimensionless. The radial equation is an equation of general
hypergeometric functions, and thus maintains the conformal
invariance. The explicit retarded Green's function is now
parametrized, in addition to the two Frolov-Thorne temperatures
also by the effective non-extremality parameter.

These results provide a starting point for further quantitative
studies of boundary conformal field theory. In particular, the
explicit form of the retarded Green's function is a key ingredient
in the calculation of the Lorentzian signature two-point and
higher point correlation functions of the boundary field theory
operators \cite{Son}.  While progress along these directions has
been made for the four-dimensional Kerr backgrounds
\cite{Becker2010}, it would be important to pursue such
calculations for the AdS black hole backgrounds, presented in this
paper.

\begin{acknowledgements}

TB would like to thank Mahmut Horta\c{c}su for various discussions
and to the Department of Physics and Astronomy at the University
of Pennsylvania for hospitality during the course of this work.
MC would like to thank Gary Gibbons, Finn Larsen and Chris Pope
for  discussions. The research of TB is supported by TUBITAK, the
Scientific and Technological Council of Turkey. MC is supported by
the DoE Grant DOE-EY-76-02- 3071, the NSF RTG DMS Grant 0636606,
the Fay R. and Eugene L. Langberg Endowed Chair, and the Slovenian
Research Agency (ARRS).
\end{acknowledgements}

\appendix*

\section{Radial equation, and small $G$  and near-extreme limit expansion}

For later purposes it turns out to be convenient to write the radial equation in terms of a dimensionless coordinate $x$ and horizon locations $x_i$, which are related to $u$ and $u_i$ as:
\begin{eqnarray}
u &=&P^{\frac{1}{3}}x, \\
u_{i} &=&P^{\frac{1}{3}}x_{i},
\end{eqnarray}
where, $P\equiv (u_{1}-u_{2})(u_{1}-u_{3})(u_{2}-u_{3})$. We note a useful
property,
\begin{equation}
(x_{1}-x_{2})(x_{1}-x_{3})(x_{2}-x_{3})=1.
\end{equation}
Furthermore, we apply another transformation:
\begin{eqnarray}
x &=&\alpha y, \\
x_{1,2} &=&\alpha y_{1,2}, \\
x_{3} &=&\frac{u_{3}}{P^{\frac{1}{3}}},
\end{eqnarray}
where $\alpha \equiv \frac{u_{1}-u_{2}}{P^{\frac{1}{3}}}$. Then the radial
equation becomes,
\begin{eqnarray}
&& \partial _{y}\{(y-y_{1})(y-y_{2})[(u_{1}-u_{2})y-u_{3}]\partial _{y}R\} \notag \\
&& +\frac{(u_{1}-u_{3})(u_{2}-u_{3})}{4}   \bigg\{ \bigg[ \frac{1}{\kappa_{1}^{2}(y-y_{1})((u_{1}-u_{2})y_{2}-u_{3})} \notag \\
&& \hspace{1.3in} -\frac{1}{\kappa_{2}^{2}(y-y_{2})((u_{1}-u_{2})y_{1}-u_{3})} \notag \\
&& \hspace{1.3in} +\frac{1}{\kappa _{3}^{2}(y_{1}-y_{2})((u_{1}-u_{2})y-u_{3})} \notag \\
&&\hspace{1.3in} +\frac{1}{G{}^{2}(u_{1}-u_{3})(u_{2}-u_{3})}\bigg] \omega {}^{2}  \notag \\
&&\hspace{0in} -\frac{c_{0}}{G(u_{1}-u_{3})(u_{2}-u_{3})}\bigg\} R=0.
\end{eqnarray}
This equation is now in a suitable form for the expansion around small $G$ as,
\begin{eqnarray}
u_{1} &\approx &u_{10}+u_{11}G+\mathcal{O}[G]^{2}, \\
u_{2} &\approx &u_{20}+u_{21}G+\mathcal{O}[G]^{2}, \\
u_{3} &\approx &-\frac{1}{G}-2m+u_{31}G+\mathcal{O}[G]^{2}, \\
\kappa _{\substack{ 1  \\ 2}}^{2} &\approx &\kappa _{\substack{ 10  \\ 20}}%
^{2}+\kappa _{\substack{ 11  \\ 21}}^{2}G+\mathcal{O}[G]^{2}, \\
\kappa _{3}^{2} &\approx &-G+\mathcal{O}[G]^{2}.
\end{eqnarray}
Now, we expand the radial function as,
\begin{equation}
R=R_{0}+\varepsilon R_{1}+\mathcal{O}[\varepsilon ]^{2}.
\end{equation}
Here, $\varepsilon \ll 1$ is a dimensionless quantity and can be chosen as,
\begin{equation}
\varepsilon =(u_{1}-u_{2})G.
\end{equation}
This may be regarded as a twofold approximation. One is $u_{1}\approx u_{2}$
(near-extremality) and the other is $G\approx 0$ (small cosmological constant)  case. The latter  one
should be valid for general values of other parameters.

The order of $\frac{1}{G^{2}}$\ term is given by\textbf{,}
\begin{equation}
\frac{1}{4}\omega ^{2}\left( 1+\frac{1}{-y_{1}+y_{2}}\right) R_{0}=0,
\end{equation}
and the $y$ values can be chosen that,
\begin{equation}
y_{1,2}=\pm \frac{1}{2},
\end{equation}
which are the values given in  \cite{CveticUW}.

The order of $\frac{1}{G}$\ term is given by,
\bea
&&\partial _{y}\left[ (y^{2}-\frac{1}{4})\partial _{y}R_{0}\right] +\frac{1}{4}%
\bigg\{ \bigg[ \frac{1}{\kappa _{10}^{2}(y-\frac{1}{2})}-\frac{1}{\kappa
_{20}^{2}(y+\frac{1}{2})} \notag \\
&& \hspace{.0in} -[2m+(u_{10}+u_{20})-(u_{10}-u_{20})y]\bigg]
\omega {}^{2} \notag \\
&& \hspace{1.3in} -c_{0}\bigg\} R_{0}=0.  \label{birbolug}
\eea
Comparing this result with the radial equation given by Cveti\v{c}-Larsen
\cite{CveticUW}, we see,
\begin{eqnarray}
\Delta _{CL} &=&(u_{10}-u_{20}) \notag \\
&=&\sqrt{[(a+b)^{2}-2(m-q)][(a-b)^{2}-2(m+q)]}, \notag
\\
M_{CL} &=&-[2m+(u_{10}+u_{20})] \notag \\
&=&-[2m-(a^{2}+b^{2}-2m)] \notag \\
&=&a^{2}+b^{2}-4m,
\end{eqnarray}
where the subscript "$CL$" denotes the related expression in the
former paper. As $G\rightarrow 0$, the radial equation reduces to
the one of \cite{CveticUW}, as expected.

In the expansion around the extreme limit  (valid now for general values of $G$ and remaining free parameters), we set,
\begin{equation}
u_{2}=u_{1}(1+\varepsilon ),
\end{equation}
where $0\leq \varepsilon \ll 1$ and $\varepsilon =0$ is the
extreme limit.

The radial function is also expanded as,
\begin{equation}
R=R_{0}+\varepsilon R_{1}+\varepsilon ^{2}R_{1}+\mathcal{O}[\varepsilon
^{3}].
\end{equation}
For the surface gravity associated with the outer horizon we have,
\begin{eqnarray}
\kappa _{1} &=&\frac{Gu_{1}^{2}[u_{1}-u_{1}(1+\varepsilon
)-u_{3}]+Gu_{1}^{2}u_{3}(1+\varepsilon )}{\sqrt{u_{1}}%
[(u_{1}+a^{2})(u_{1}+b^{2})+abq]}  \notag \\
&=&\frac{\varepsilon Gu_{1}^{2}(u_{3}-u_{1})}{\sqrt{u_{1}}%
[(u_{1}+a^{2})(u_{1}+b^{2})+abq]} \\
&\equiv&\varepsilon \kappa _{11},  \notag
\end{eqnarray}
the surface gravity associated with the inner horizon is,
\begin{eqnarray}
\kappa _{2} &=&\frac{Gu_{1}^{2}(1+\varepsilon
)^{2}(\varepsilon
u_{1}-u_{3})+Gu_{1}^{2}(1+\varepsilon )u_{3}}{\sqrt{u_{1}(1+\varepsilon )%
}\{[u_{1}(1+\varepsilon )+a^{2}][u_{1}(1+\varepsilon
)+b^{2}]+abq\}}  \notag
\\
&\equiv&-\varepsilon \kappa _{11}+\varepsilon ^{2}\kappa _{22}+\mathcal{O}%
[\varepsilon ^{3}],
\end{eqnarray}
and for the surface gravity associated with the third horizon we
have,
\begin{eqnarray}
\kappa _{3} &=&\frac{Gu_{3}^{2}[u_{3}-u_{1}(2+\varepsilon
)]+Gu_{1}^{2}u_{3}(1+\varepsilon )}{\sqrt{u_{3}}%
[(u_{3}+a^{2})(u_{3}+b^{2})+abq]}  \notag \\
&\equiv&\kappa _{30}+\varepsilon \kappa _{21}.
\end{eqnarray}
When we plug them into the radial equation, the $\varepsilon ^{0}$ order
gives an equation for $R_{0}$ only,
\begin{widetext}
\begin{equation}
\frac{d}{du}\left[ G(u-u_{1})^{2}(u-u_{3})\frac{dR_{0}}{du}\right] +\frac{1}{%
4}\left\{ \left[ \frac{Gu_{1}(\kappa _{11}u_{1}(u-u_{3})+2\kappa
_{22}(u-u_{1})(u_{1}-u_{3}))}{\kappa _{11}^{3}(u-u_{1})^{2}}+\frac{%
G(u_{1}-u_{3})^{2}}{\kappa _{30}^{2}(u-u_{3})}+\frac{1}{G}\right] \omega
^{2}-c_{0}\right\} R_{0}=0,  \label{sifirmert}
\end{equation}
and it is solved in terms of confluent Heun's functions ($H_{C}$),
\begin{eqnarray}
R_{0} &=&e{^{-{\frac{\,i\omega u_{1}}{2\kappa _{11}\,\left( u-u_{1}\right) }}%
}}\left\{ C_{1}\,\left( u-u_{3}\right) ^{{\frac{i\omega }{2\kappa _{30}}}%
}\left( u-u_{1}\right) ^{-{\frac{4\,\kappa _{30}+i\omega }{2\kappa _{30}}}%
}\right.  \notag \\
&&\times H_{C}\left[ {\frac{-i\omega u_{1}}{\kappa _{11}\,\left(
u_{1}-u_{3}\right) }},{\frac{i\omega }{\kappa _{30}}},2,{\frac{\omega
^{2}u_{1}\,\kappa _{22}}{2\left( u_{1}-u_{3}\right) \kappa _{11}^{3}}},{%
\frac{[4\left( u_{1}-u_{3}\right) {G}^{2}-c_{0}G+\omega ^{2}]\kappa
_{11}^{3}-2\,u_{1}\,\omega ^{2}{G}^{2}\kappa _{22}}{4\left(
u_{1}-u_{3}\right) {G}^{2}\kappa _{11}^{3}}},{\frac{u-u_{3}}{u-u_{1}}}\right]
\notag \\
&&+C_{2}\left( u-u_{3}\right) ^{-{\frac{\,i\omega }{2\kappa _{30}}}}\,\left(
u-u_{1}\right) ^{{\frac{-4\,\kappa _{30}+i\omega }{2\kappa _{30}}}}
\\
&&\left. \times H_{C}\left[ {\frac{-i\omega u_{1}}{\kappa _{11}\,\left(
u_{1}-u_{3}\right) }},{\frac{-i\omega }{\kappa _{30}}},2,{\frac{\omega
^{2}u_{1}\,\kappa _{22}}{2\left( u_{1}-u_{3}\right) \kappa _{11}^{3}}},{%
\frac{[4(\,u_{1}-u_{3}){G}^{2}-c_{0}G+\omega ^{2}]\kappa
_{11}^{3}-2\,u_{1}\,\omega ^{2}{G}^{2}\kappa _{22}}{4\left(
u_{1}-u_{3}\right) {G}^{2}\kappa _{11}^{3}}},{\frac{u-u_{3}}{u-u_{1}}}\right]
\right\} , \notag
\end{eqnarray}
\end{widetext}
where $C_{1}$ and $C_{2}$ are constants.

In order to identify the near-horizon behavior of the radial
equation in the near-extreme limit we will apply the following
parametrization:
\begin{eqnarray}
\omega &\rightarrow &\eta \omega _{h}, \\
u &\rightarrow &u_{1}(1+\eta \rho)^2,
\end{eqnarray}
and take the limit $\eta \rightarrow 0$. Then, for the leading order of $%
\eta $, Eq. (\ref{sifirmert}) can be solved in terms of Bessel's
functions,
\bea
R_{0}&=&\frac{1}{\sqrt{\rho}}\bigg[ C_{1}^{\prime }\,J{\left( -\frac{1}{2}\sqrt{1+%
{\frac{c_{0}}{G\left( u_{1}-u_{3}\right) }}},\,{\frac{\omega _{h}}{%
4\kappa _{11}\,\rho}}\right) }  \notag \\
&&+C_{2}^{\prime }\,Y{\left( -\frac{1}{2}\sqrt{1+{%
\frac{c_{0}}{G\left( u_{1}-u_{3}\right) }}},{\frac{\omega
_{h}}{4\kappa _{11}\,\rho}}\right) }\bigg] {,}  \label{kasim0}
\eea
where $C_{1}^{\prime }$ and $C_{2}^{\prime }$ are constants and
$c_{0}$ is the separation constant. This solution formally agrees
with the solution of the extreme case as the Whittaker functions
are transformed into Bessel's functions as $m_{1}=m_{2}=0$.  Since
in the near horizon regime the equation is a hypergeometric
equation, it possesses the  $SL(2,{\bf R})^2$ conformal symmetry.

\end{document}